# Circuit models for Sierpinski gasket antennas

Walter Arrighetti, Peter De Cupis, Giorgio Gerosa

***Abstract*—A lumped-parameter impedor-oriented and a 2-port-network-oriented circuit models for the Sierpinski gasket prefractal antenna are presented. With the former, the voltage and current patterns give a detailed understanding of the electromagnetic fields' self-similar distribution throughout the antenna geometry; on the other hand model complexity exponentially increases with the prefractal iteration order. The latter "black-box" model only controls port-oriented global parameters which are the ones commonly used in antennas' circuit models and its complexity is independent of prefractal order. The "black-box" model is also shown to converge, at fractal limit, to a reciprocal triangular network.**

***Index Terms*—circuit modeling, current distribution, fractal, iterated function system (IFS), iterative methods, Sierpinski gasket, two-port networks.**

## I. Introduction

FRACTAL antennas have been studied, built and commercialized for a considerable while. The fractal geometry (i.e. the replication of the same set over and over on infinitely decreasing length scales [1], [2]) of a cavity or radiating element as well is responsible for the self-similarity of its spectrum, in both the continuous [3]–[5] and discrete cases [6]. For example, properly synthesized fractal antennas feature multi-band properties, as several theoretical and experimental studies confirmed [7]–[14]. In particular, some devices for applications in modern mobile radio-communication systems based on Sierpinski gasket-like geometries have a log-periodic behaviour as far as both input parameters and radiation patterns are concerned.

Moreover, the intrinsic planar nature of the geometry provides an efficient implementation by means of 2D self-resonant wire monopoles, that can be printed on a fibreglass substrate through standard techniques used for the manufacturing of printed electronic boards' circuits.

Fractal geometries used in electromagnetic applications are usually generated via Iterated Function Systems (*IFS*s), i.e. the infinite recursion of a contraction mappings' sequence to an initiator domain [1]. The limit fractal set cannot be used to model real structures as such infinitely small length-scales are neither physically plausible nor practically realizable. Real structures are produced by stopping the IFSs' iteration up to a finite step and the obtained domain is thus called a *pre*fractal.

Authors (email: arrighetti, decupis, gerosa@die.uniroma1.it) are with the Department of Electronic Engineering, Università degli Studi di Roma "La Sapienza", 00184 Rome, Italy, www.die.uniroma1.it/strutture/labcem/.

Two network-based methods for the analysis of electromagnetic Sierpinski gasket-like structures will be presented here: a lumped-parameters one (see Chapter *III*) and a "black-box" one (see Chapter *IV*). They are presented as depending on the prefractal order of the structure (i.e. the step which the IFS was stopped at), such that the asymptotic behaviour of electric parameter, as long as the prefractal tends to the limit fractal set, are highlighted and made comparable with efficiency and production costs.

## II. From Self-similar Graphs to Electric Circuits

### A. Self-similar graphs

The graph-analogue to an IFS (but without any 'metric' information) is a mapping $T\colon G\to G^p$, where $G=(N,L)$ is a finite or countable, simple graph (for the sake of the electric theory discussed from Subsection *II.C* on, it can also be non-simple), and $G^p=(\oplus^p N, L_p)$. For basic graph definitions see Appendix.

The node set of $G^p$ is simply $\oplus^p N := N\oplus N\oplus\ldots\oplus N$ ($p$ times); this stands for the nodes of $G^p$ being $p$ "copies" of $G$'s ones. The set $L_p$ is, by definition of graph, a subset of the set $\oplus^p N\times\oplus^p N$, but it can also be a superset of the $L^p$ (cartesian product of $L$ by itself $p$ times) so, in general, $L^p\subseteq L_p\subseteq\oplus^p N\times\oplus^p N$: this reflects the fact that each "copy" of the original graph is intra-connected the same way, but map $T$ *might* add links between nodes of different "copies" (usually determining the connectedness of graphs $G^p$, for any $p$, if $G$ is connected as well).

Let $G_0=(N,L)$ be such a finite graph (oriented, for generality's sake), let $G_n=T^n(G_0)$ $\forall n\in\mathbb{N}$, and $H_0\in\mathbb{M}_{\mathrm{Card}N}(\mathbb{Z}_3)$ its skew-symmetric incidence matrix (hereinafter let '$\mathrm{Card}A$' be the number of elements of set $A$). Whatever is the chosen ordering of the nodes, provided that the nodes of each copy either "follow" or "precede" those of the other copies (and their inner ordering is the same of $G_0$), the incidence matrix $H_n$ of the $n^{\text{th}}$ iteration $G_n$, as shown in [15]–[17], is computed as:

$$H_n = V_n + \bigoplus^p H_{n-1}\,, \quad \forall n\in\mathbb{N}. \tag{1}$$

Thus $H_n$ is a quasi-diagonal-block matrix, whose $p$ diagonal blocks are all equal to the incidence matrix of the $(n-1)^{\text{th}}$ iterate, which a sparse matrix $V_n$ is added to. Latter matrix is nonzero only if the copies of each graph get some interconnections under $T$ (i.e. the $G_1$'s link-set is larger than just $p$ copies of $G_0$'s link-set). If $G_0$ has a finite number of nodes, then $G_n$ has $p^n \mathrm{Card}\, G_0$ nodes, which is the number of columns and rows of $H_n$ and $V_n$ as well. The number of nonzero elements in $V_n$ is twice the number of inter-connections between the copies of the original graph for every new iteration[1].

The infinite graph obtained by infinitely composing map $T$ with itself is called the *self-similar graph* of $T$, since its incidence relation is similar under all of its subcopies' hierarchies and is formally indicated as a 'limit graph':

$$G_\infty := \lim_n G_n = \lim_n T^n(G_0)\,. \tag{2}$$

The self-similar structure of $G_n$ graphs ($\forall n \in \mathbb{N}$) is thus reflected in the multiscale structure of their incidence matrices, as the $V_n$ matrices have a 'dilated' sparseness depending on $n$, and the diagonal blocks are all copies of the previous-iteration matrices. The finite iterations of $T$ are usually called *prefractal graph*s, in analogy with IFSs.

### B. *The Sierpinski gasket graph*

The *Sierpinski gasket graph* is any limit graph obtained by recursively copying three times ($p$=3) any simple 3rd-order graph and adding three links among these copies [15]. As initiator, a complete 3rd-order graph, i.e. a *triangle* graph, is considered, and the action of the map is shown in Fig. 1.

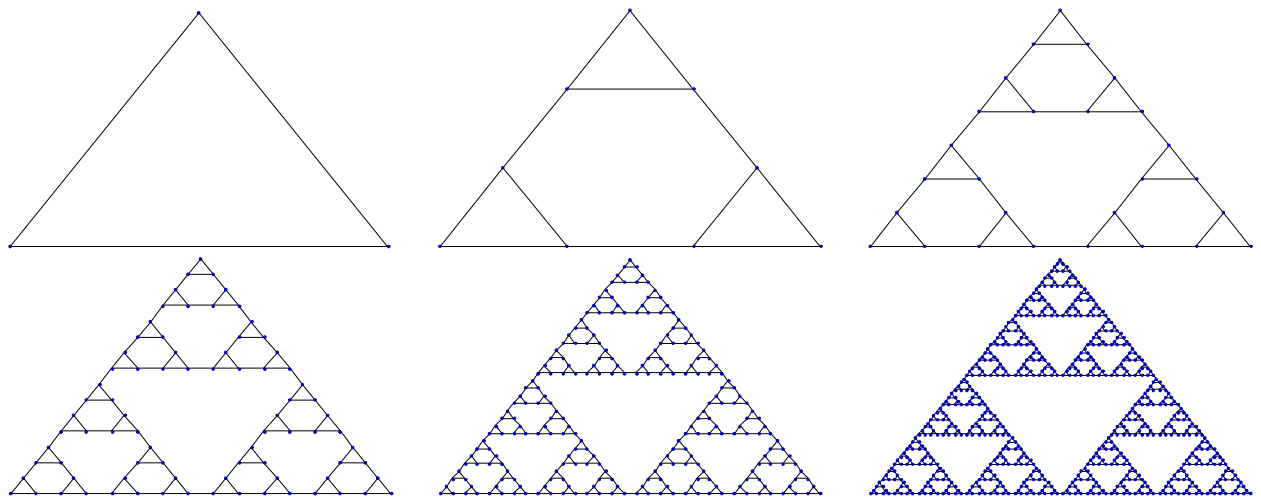

Fig. 1. Sierpinski gasket prefractal graphs $\check{S}_n$, $0 \le n \le 5$.

A smart algorithmic nodes' ordering method consists in choosing either one of the two orientations for the original triangle graph $\check{S}_0$ (e.g. the clockwise one) and assigning one *ternary* (base-3) digit to each of the nodes: $0_3$, $1_3$ and $2_3$. Then every node of the $n^{\text{th}}$-iteration graph $\check{S}_n$ is inductively labeled with an $(n+1)$-digit ternary integer, whose $n$ less significant digits reflect the node ordering of its main 3 copies (i.e. according to the inner ordering of $\check{S}_{n-1}$) and the most significant digit orders the three copies themselves (e.g. clockwise).

$\check{S}_1$ is thus made up of three copies of $\check{S}_0$: the nodes of the 'first' copy are labeled $00_3$=0, $01_3$=1 and $02_3$=2; nodes of the 'second' copy $10_3$=3, $11_3$=4 and $12_3$=5; nodes of the 'last' copy $20_3$=6, $21_3$=7 and $22_3$=8; and so on for $\check{S}_n$, $\forall n \in \mathbb{N}$, which has $3^{n+1}$ nodes.

According to the formalism given in Subsection *II.A*, the (non-invertible) incidence matrices $H_n$ of the Sierpinski gasket's prefractal graphs $\check{S}_n$ are easily computed[2] thanks to this ordering as[3]:

$$H_0 = \begin{pmatrix} 0 & 1 & -1 \\ -1 & 0 & 1 \\ 1 & -1 & 0 \end{pmatrix}; \tag{3}$$

$$V_n = \left( \delta_{i,\left\lceil \frac{3^n}{2} \right\rceil} \delta_{j,3^n+1} - \delta_{i,3^n} \delta_{j,2\cdot 3^n+1} + \delta_{i,2\cdot 3^n} \delta_{j,\frac{5\cdot 3^n-1}{2}+1} \right)_{1 \le i < j \le 3^{n+1}}$$

Next two 9×9 and 27×27 incidence matrices are shown for their multiscale structure ('*' represents any skew-symmetric blocks, bigger '0's any zero-blocks):

$$H_1 = \left( \begin{array}{ccc|ccc|ccc} 0 & 1 & -1 & 0 & 0 & 0 & 0 & 0 & 0 \\ -1 & 0 & 1 & 1 & 0 & 0 & 0 & 0 & 0 \\ 1 & -1 & 0 & 0 & 0 & 0 & -1 & 0 & 0 \\ \hline & & & 0 & 1 & -1 & 0 & 0 & 0 \\ & * & & -1 & 0 & 1 & 0 & 0 & 0 \\ & & & 1 & -1 & 0 & 0 & 1 & 0 \\ \hline & & & & & & 0 & 1 & -1 \\ & * & & & * & & -1 & 0 & 1 \\ & & & & & & 1 & -1 & 0 \end{array} \right);$$

$$H_2 = \left( \begin{array}{ccccccccc|ccccccccc|ccccccccc}
0&1&-1&0&0&0&0&0&0&&&&&&&&&&&&&&&&&& \\
-1&0&1&1&0&0&0&0&0&&0&&&0&&&0&&&0&&&0&&&0& \\
1&-1&0&0&0&0&-1&0&0&&&&&&&&&&&&&&&&&& \\
&&&0&1&-1&0&0&0&0&0&0&&&&&&&&&&&&&&& \\
&*&&-1&0&1&0&0&0&1&0&0&&0&&&0&&&0&&&0&&&0& \\
&&&1&-1&0&0&1&0&0&0&0&&&&&&&&&&&&&&& \\
&&&&&&0&1&-1&&&&&&&&&&0&0&0&&&&&& \\
&&&&*&&-1&0&1&&0&&&0&&&0&&0&0&0&&0&&&0& \\
&&&&&&1&-1&0&&&&&&&&&&-1&0&0&&&&&& \\
\hline
&&&&&&&&&0&1&-1&0&0&0&0&0&0&&&&&&&&& \\
&*&&&&&&*&&-1&0&1&1&0&0&0&0&0&&0&&&0&&&0& \\
&&&&&&&&&1&-1&0&0&0&0&-1&0&0&&&&&&&&& \\
&&&&&&&&&&&&0&1&-1&0&0&0&&&&&&&&& \\
&&&&*&&&&&&*&&-1&0&1&0&0&0&&0&&&0&&&0& \\
&&&&&&&&&&&&1&-1&0&0&1&0&&&&&&&&& \\
&&&&&&&&&&&&&&&0&1&-1&&&&0&0&0&&& \\
&*&&&&&&*&&&&&&*&&-1&0&1&&0&&0&0&0&&0& \\
&&&&&&&&&&&&&&&1&-1&0&&&&0&1&0&&& \\
\hline
&&&&&&&&&&&&&&&&&&0&1&-1&0&0&0&0&0&0 \\
&&&&*&&&&&&*&&&&&&*&&-1&0&1&1&0&0&0&0&0 \\
&&&&&&&&&&&&&&&&&&1&-1&0&0&0&0&-1&0&0 \\
&&&&&&&&&&&&&&&&&&&&&0&1&-1&0&0&0 \\
&*&&&&&&*&&&&&&*&&&&&&*&&-1&0&1&0&0&0 \\
&&&&&&&&&&&&&&&&&&&&&1&-1&0&0&1&0 \\
&&&&&&&&&&&&&&&&&&&&&&&&0&1&-1 \\
&&&&*&&&&&&*&&&&&&*&&&&&&*&&-1&0&1 \\
&&&&&&&&&&&&&&&&&&&&&&&&1&-1&0
\end{array} \right).$$

To complete the description of such graphs, incidence matrices are singular $\forall n \in \mathbb{N}_0$ and the links of $\check{S}_n$ are counted to be $3(3^{n+1}-1)/2$ (so the graph is totally connected).

### C. *From incidence to admittances*

This paragraph only summarizes the basics of a well known circuit theorists' method to solve lumped-parameters circuit equations called the *nodes' method* [18] and its strong linking with graph theory. It can also be extended to distributed-parameters circuits. As long as electrically linear impedor elements alone are involved, such a network is represented by a finite ($N$ nodes and $L$ links) non-ordered graph with every link associated to an impedor, whose Laplace-domain admittance $y(s)$, $s \in \mathbb{C}$, is the transfer function between its voltage $V(s)$ and current $I(s)$. The graph has $P \in \mathbb{N}$ connected components, but electrical constraints are imposed between.

Once a reference node is chosen for the entire graph (*ground node*) its degrees of freedom are the $N-P$ *node*

[1] *Twice* because $V_n$ must be skew-symmetric.

[2] The closed form for $V_n$ only refers to its *upper* triangle, as the indexing format $1 \le i < j \le 3^{n+1}$ defines its skew-symmetry. $\delta_{i,j}$ is the Kronecker delta.

[3] Symbol '⌈·⌉' denotes the *ceiling* function, e.g.: $\lceil e \rceil = 3$, $\lceil 0.25 \rceil = 1$, $\lceil -\pi \rceil = -3$.

*voltages* $\mathbf{V}(s)$ and the $L{-}N{+}P$ *link currents* $\mathbf{I}(s)$ on each connected part's fundamental trees (cfr. Appendix). Any other voltages and currents depend on these two vectors, being a canonical pair of Lagrangian coordinates [19] for the circuit.

Nodes' method consists in bulding the *symmetric admittances' matrix* $Y(s)$ whose $k^{\text{th}}$ diagonal element ($1{\le}k{\le}N{-}P$) is the sum of the admittances on all the $k^{\text{th}}$-node-related links, and the $(i,j)^{\text{th}}$ triangular element ($i{\neq}j$) is the sign-reversed sum of the admittances between $i^{\text{th}}$ and $j^{\text{th}}$ node. Letting $\mathbf{I}_{\mathrm{g}}(s)$ be the vector whose $k^{\text{th}}$ element is the sum of all the impressed (independent-generator-driven) currents oriented *toward* the $k^{\text{th}}$ node, lagrangian voltages are computed solving the linear system of equations:

$$Y(s)\mathbf{V}(s) = \mathbf{I}_{\mathrm{g}}(s)\,. \tag{4}$$

Symbolically, admittances' and incidence matrices are related to each other: they both have the same zero elements (presence of an impedor is equivalent to incidence of its two nodes), but the former is symmetric, the latter is skew-symmetric. Diagonal elements are 0 because *cycle*s (links of a node to itself) are not allowed in such graphs.

## III. Endoscopic model of the Sierpinski gasket

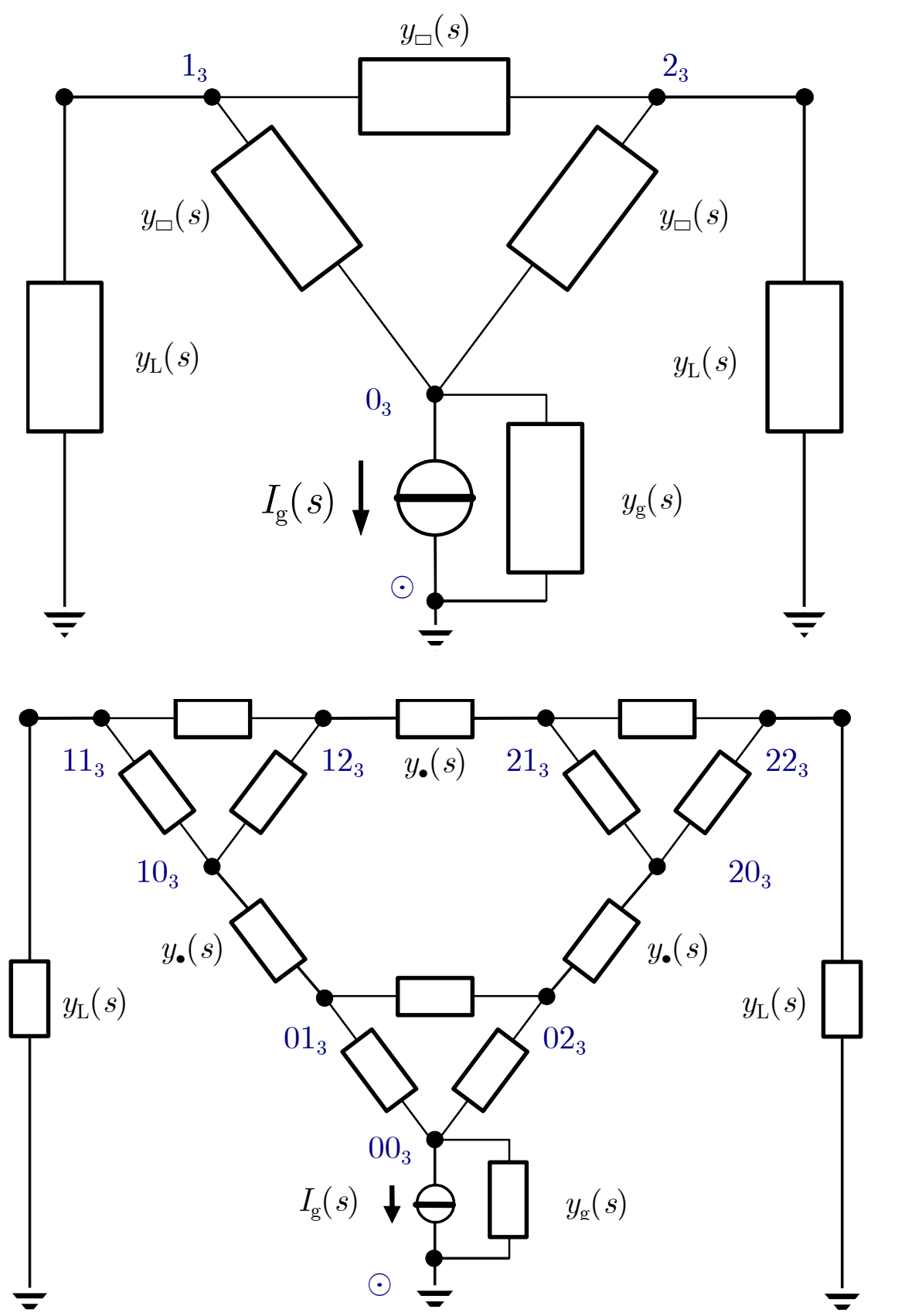

Fig. 2. Initiator (top) and first iteration (bottom) of the Sierpinski gasket electric circuit. Node labels, reference node, impedor admittances and current generator are shown.

### A. *Putting all the impedors together*

Let $\check{S}_n$ be the Sierpinski gasket's $n^{\text{th}}$-iteration graph ($n{\in}\mathbb{N}_0$), as defined in Subsection *II.B*. The modeled electric network is recursively built starting from $\check{S}_0$, which is a triangular circuit whose 3 links have the same impedor admittance $y_\square(s)$, then adding appropriate feeding impedors (read below and cfr. Fig. 2). The interconnection links between the previous iteration's 3 copies all have $y_\bullet(s)$ impedor admittance.

$\check{S}_n$ has, $\forall n{\in}\mathbb{N}_0$, 3 *distal* nodes (its geometrically, but not path-wise, farthest vertices). As shown in Fig. 2 an independent outward-oriented current generator $I_{\mathrm{g}}(s)$, together with its parallel admittance $y_{\mathrm{g}}(s)$, is attached to distal node $00\ldots0_3{=}0$; their other end is grounded and chosen as reference node (indicated as '$\odot$' in Fig. 2, cfr. Subsection *II.C*). Distal nodes $11\ldots1_3{=}(3^{n+1}{-}1)/2$ and $22\ldots2_3{=}3^{n+1}{-}1$ are both attached to load admittances $y_{\mathrm{L}}(s)$, and whence grounded [20].

The admittances' matrix $Y_n(s)$ of the electric circuit $\check{S}_n$ is computed according to Subsections *II.A*,*C* as[4]:

$$Y_n(s) = Y_n^{\mathrm{o}}(s) + L_n(s)\,, \qquad \forall n{\in}\mathbb{N}_0; \tag{5}$$

$$Y_0^{\mathrm{o}}(s) = y_\square(s)\begin{pmatrix} 2 & -1 & -1 \\ -1 & 2 & -1 \\ -1 & -1 & 2 \end{pmatrix};$$

$$L_n(s) = \operatorname{diag}\big(y_{\mathrm{g}}(s)\delta_{i,1} + y_{\mathrm{L}}(s)(\delta_{i,3^n+1} + \delta_{i,2\cdot 3^n+1})\big)_{1\le i\le 3^{n+1}}\,;$$

$$Y_n^{\mathrm{o}}(s) = V_n(s) + \bigoplus^{3} Y_{n-1}^{\mathrm{o}}(s)\,;$$

$$V_n(s) = -y_\bullet(s)\cdot$$

$$\cdot\Big(\delta_{i,\lceil\frac{3^n}{2}\rceil}\delta_{j,3^n+1} + \delta_{i,3^n}\delta_{j,2\cdot 3^n+1} + \delta_{i,2\cdot 3^n}\delta_{j,\frac{5\cdot 3^n-1}{2}+1}\Big)_{1\le i\le j\le 3^{n+1}}\,.$$

$V_n$ matrices stand for interconnection impedors, whereas $L_n$ matrices stand for grounding ones. The impressed currents' vector has one generator $I_{\mathrm{g}}(s)$ as first and non-zero element:

$$\mathbf{I}_{\mathrm{g},n}(s) = -I_{\mathrm{g}}(s)\big(\delta_{k,1}\big)_{1\le k\le 3^{n+1}}\,. \tag{6}$$

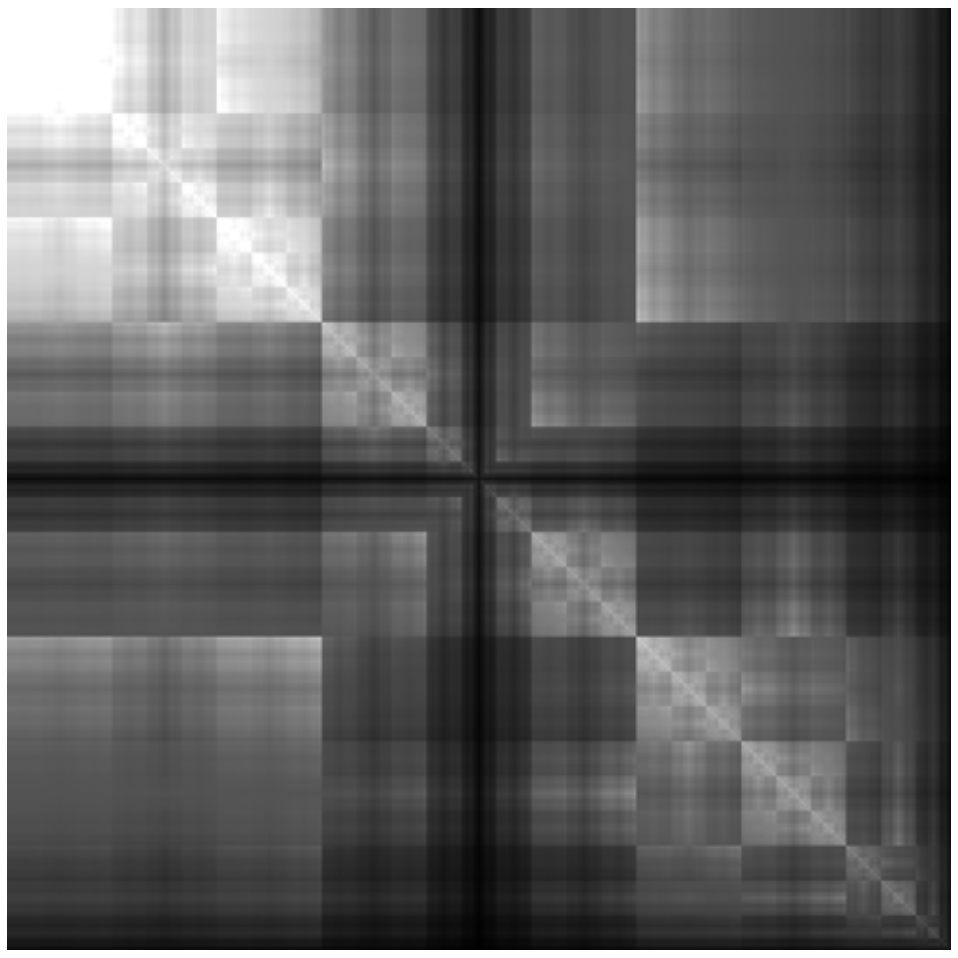

Fig. 3. Representation of the inverse of the homogeneous $9^{\text{th}}$-iteration Sierpinski gasket's admittances' $3^{10}{\times}3^{10}$ matrix $Y_9^{-1}$.

[4] Note that $V_n$ is indexed as $1{\le}i{\le}j{\le}3^{n+1}$ due to its *symmetry*; cfr. *II.B*.

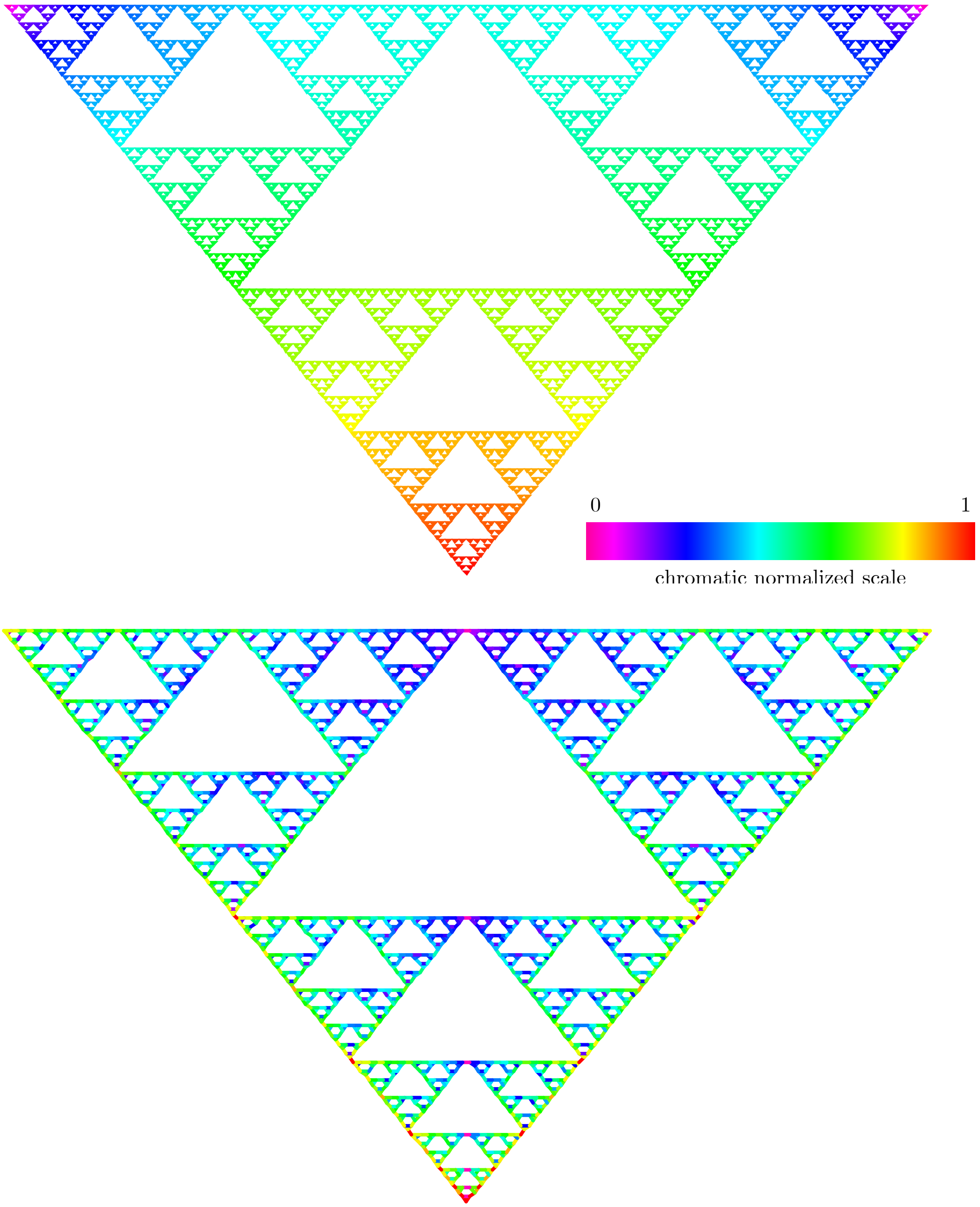

Fig. 4. Self-similar distributions of node voltages (top) and link currents (bottom) for the homogeneous 9th-iteration Sierpinski gasket.

Other generators in parallel to any impedors could be added. This is the simplest generator configuration to model a Sierpinski gasket antenna fed from just the "lower" vertex.

The $3^{n+1}$ node voltages, represented by vector $\mathbf{V}_n(s)$, are given by the identity

$$\mathbf{V}_n(s) = Y_n^{-1}(s)\mathbf{I}_{\mathrm{g},n}(s)\,, \tag{7}$$

i.e. by $-I_{\mathrm{g}}(s)$ times the first column of $Y_n(s)$'s inverse.

### A. *Homogeneous Sierpinski gasket circuit*

Geometric graph self-similarity produces a self-similar distribution of electric quantities throughout the circuit, even in the simple case of a permanent-regime feeding and modeling $\check{S}_n$ as a homogeneous network, i.e. using identical impedors $y_{\square}(s)\equiv y_{\bullet}(s)\equiv y_{\mathrm{L}}(s)$. In this case there is a complete proportionality between node voltages and the impedors' impedances. The simplest model employs a purely resistive network, $y_{\square}(s)=y_{\bullet}(s)=y_{\mathrm{L}}(s)\equiv R^{-1}>0$, an ideal generator $y_{\mathrm{g}}(s)\equiv\infty$ and constant (DC) feeding $I_{\mathrm{g}}(s)\equiv I_0$. The linear system to solve is thus purely numerical,

$$Y_n\mathbf{V}_n = \mathbf{I}_{\mathrm{g},n}\,, \tag{8}$$

where $Y_n \in \mathbb{M}_{3^{n+1}}\left(R^{-1}\mathbb{Z}\right)$ and $\mathbf{I}_{\mathrm{g},n}=(-I_0,0,0,\ldots,0)$.

Self-similarity is well reflected in the admittances' matrix and in its inverse as well, which is represented in Fig. 3 as a grayscale image whose dots' brightness is proportional to the magnitude of the matrix elements for $n$=9.

A self-similar block structure in the matrix is still observed: different scaling between diagonal and non-diagonal blocks is present, each one associated to the electric behaviour of every subnetworks' hierarchies in the Sierpinski gasket circuit. This result, obtained with the nodes' method shown in Chapter *II* and based on just a few general assumptions about the circuit's electrical conditions (in DC regime), was first observed in 2001 [15], and the same structure of the resolvent matrix was found using a new discrete-topology method for field equations called Topological Calculus [5], [15], [21]. It is also nonetheless coherent with numerical results obtained in [22] with a full-wave method of moments (*MoM*) simulation of a prefractal Sierpinski gasket antenna.

The self-similar distribution of such voltages $\mathbf{V}_n$ and (by simple constraints on the ideal resistors of the circuit) of link currents $\mathbf{I}_n$ are shown in Fig. 4 for the case $n$=9. Direct inspection of latter figure easily shows that the currents propagate with the same diffusion pattern along each of the triangular sub-networks' 9 hierarchies and, by Babinet's Theorem, are zero at the central vertical symmetry axis of the circuit (triangle height of the $00\ldots0_3$-labeled triangle vertex). Currents' distribution is also in accordance with numerical results obtained via a scattering-matrix transmission-line approach to the Sierpinski gasket antenna [23].

### *B. Results of the resonant Sierpinski circuit*

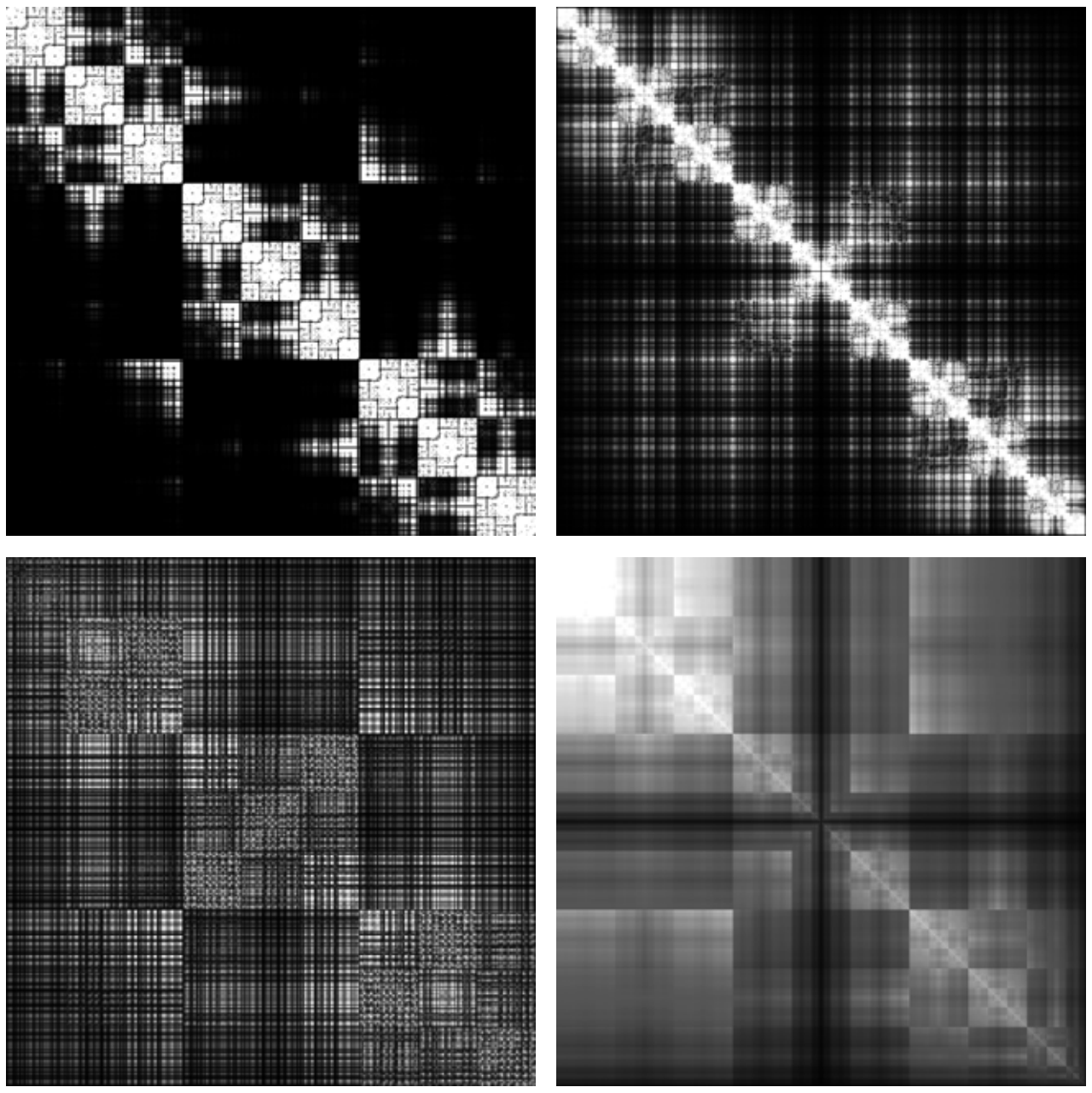

Fig. 5. Modulus inverse of 9th-iteration Sierpinski gasket's admittances' matrix $\left|Y_9^{-1}(i\omega)\right|$ near the resonance $\omega_{\text{res}}$ (left to right, top to bottom): $|\nu-\nu_{\text{res}}| = 300,\ 100,\ 30,\ 0$ Hz.

A frequency-domain study of the Sierpinski gasket antenna is carried out using reactive impedors too. For instance, the dependence on feeding frequency can be studied assuming that the $y_\square(s)$ impedors, associated to each of the antenna's triangular patches, are purely dissipative elements (resistors of resistance $R$>0) while the $y_\bullet(s)$ impedors, modeling the interconnections, are LC-parallels resonating at $\omega_{\text{res}} = 1\big/\sqrt{LC} > 0$ :

$$y_{\text{L}}(s) = y_\square(s) \equiv \frac{1}{R}; \qquad y_\bullet(s) = \frac{1+s^2LC}{sL}. \tag{9}$$

Used test values are $R$=1 μΩ, $L$=0.1 μH and $C$=1 nF (thus $\omega_{\text{res}}$=$10^8$ rad, $\nu_{\text{res}}$≃15.91 MHz). An ideal generator is still considered, so $y_{\text{g}}(s)$≡∞.

The admittances' matrices are shown in Fig. 5 for different, almost-resonating values of frequency $\nu$; dots' brightness displays the modulus of the various matrix elements..

The resonating circuit, i.e. at $\nu = \nu_{\text{res}}$, formally behaves like the homogeneous model introduced in Subsection *III.A*, but different, self-similar currents' distribution is observed for different values as well.

## IV. Black-Box model of the Sierpinski gasket

### *A. Self-similar N-port networks*

An alternative method for circuit models of self-similar structures is here proposed. As opposed to the one in Subsections *III.A*,*B*, this method uses multi-port networks as initiators, whose global electric parameters are known (e.g. *impedance matrix*) instead of the fine structure inside the electric circuit itself; that is why it is a somewhat "black-box" model.

Let $Z_0(s) \in \mathbb{M}_N\left(\mathbb{R}^{\mathbb{Q}}[s]\right)$ be the impedance matrix[5] of an *N-port network* and $Z_n(s) = TZ_{n-1}(s) \equiv T^nZ_0(s)$, where $T : \mathbb{M}_N\left(\mathbb{R}^{\mathbb{Q}}[s]\right) \to \mathbb{M}_N\left(\mathbb{R}^{\mathbb{Q}}[s]\right)$ maps $Z_0(s)$ to an *N*-port impedance matrix (hereinafter explicit indication of the Laplace's variable $s$∈ℂ will be suppressed). This means that the matrix transformation, no matter how complex it is, must always produce physically meaningful impedance matrices.

Map $T$ models the construction of a new *N*-port network starting from the initiator network (or more copies of it, as is the case of self-similar networks).

Once a suitable *complete* function space is chosen for such matrices, conditions for map $T$ such that the iteration procedure still converges to a fixed *attractor* *N*-port network for a subset of initial *N*-port networks (i.e. a *basin of attraction* in the dynamical systems' language) can be imposed.

The existence of an attractor *N*-port network, i.e. a limit

[5] Any other matrices could be considered in place of it (when they exist), as either the *scattering*, the *transmission* and the *admittance matrix* for example (the latter not to be confused with the admittances' matrix defined in Subsection *II.C*). $\mathbb{R}^{\mathbb{Q}}[s]$ is the space of real algebraic functions of variable $s$∈ℂ (although the elements of such a matrix have further restrictions depending on the inner electrical network structure).

impedance matrix

$$Z_\infty(s) = \lim_{n\to\infty} Z_n(s) \equiv \lim_{n\to\infty} T^n Z_0(s)$$

still having the required properties, consists in proving $T$ to be a contraction in the chosen complete function space (defining the above limit operator as well) [2].

Map $T$ usually "entangles" the elements of the matrix iterates, leading them to have a self-similar functional structure which, if a limit matrix exists, must converge to *stable* formulas under the choice of the initiator matrix $Z_0(s)$.

In this sense such a map setting can be regarded as the multi-port-network equivalent of an IFS; the limit network and those obtained under any finite iterations are called *self-similar* and *prefractal N-port network*s, respectively.

A similar approach for the computing of the scattering matrix of a triangular initiator transmission-line network was proposed [23]: the network was built with a recursive paradigm analogue to the Sierpinski one shown in Figs. 6–7 and introduced below in Subsection *IV.B* for the impedance matrix instead. A network-based iterative model for a Sierpinski *carpet*'s impedance matrix was also proposed [24].

### B. *Iterative construction of the Sierpinski 2-port network*

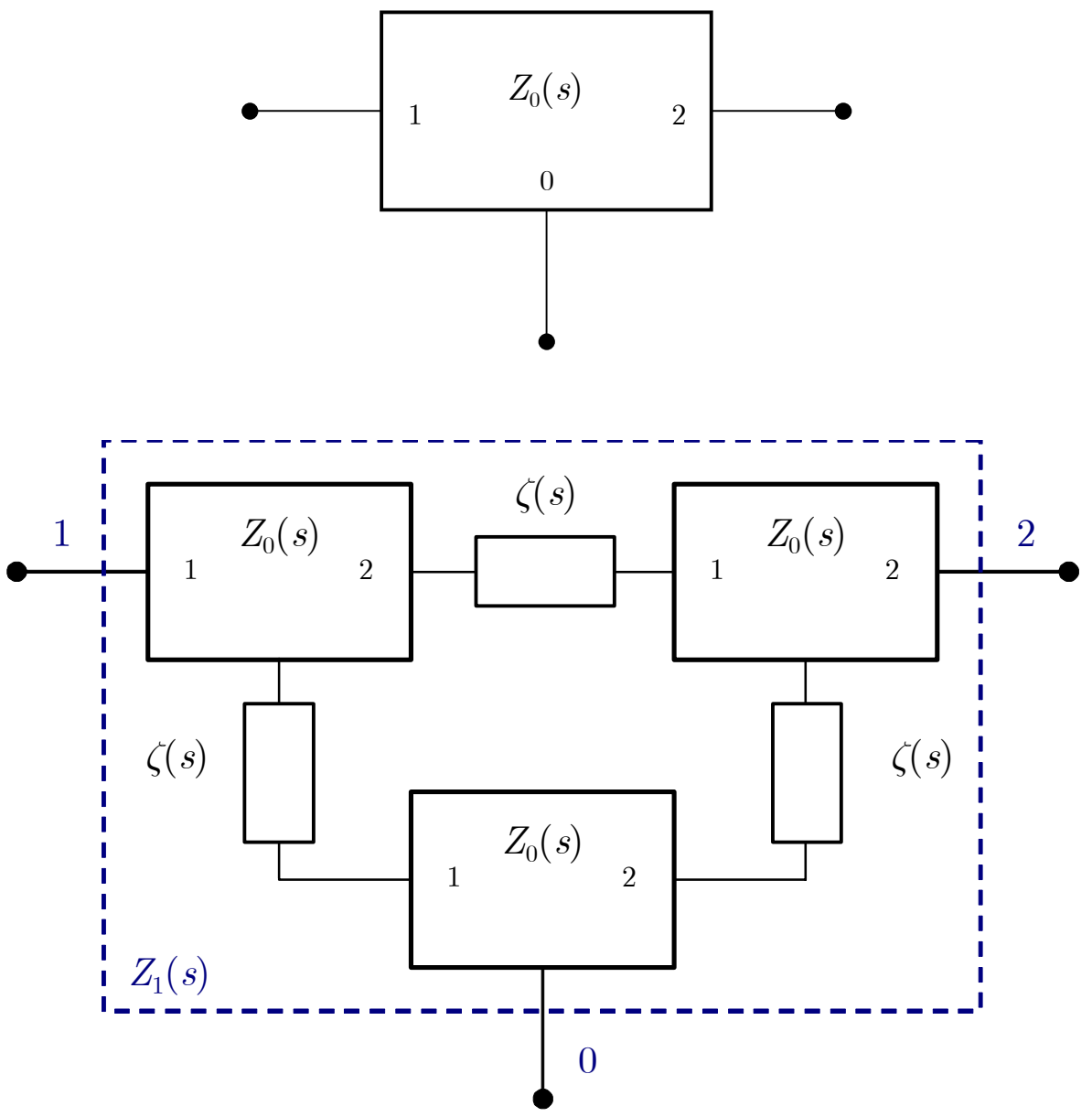

Fig. 6. Initiator $Z_0$ (top) and first step $Z_1 = \check{s}_\zeta(Z_0)$ (bottom) in the construction of the Sierpinski 2-port network.

The Sierpinski network model is carried out by a one-parameter $2\times 2$ impedance matrix map $\check{s}_\zeta : \mathbb{M}_2(\mathbb{C}) \to \mathbb{M}_2(\mathbb{C})$, $\forall \zeta \in \mathbb{C}$:

$$Z = \begin{pmatrix} z_{1,1} & z_{1,2} \\ z_{2,1} & z_{2,2} \end{pmatrix} \mapsto Z + \operatorname{diag}\big(z_{1,1} + \zeta, z_{2,2} + \zeta\big) + \\ + \frac{\big((2\delta_{i,j} - 1)(z_{i,i} + \zeta)(z_{j,j} + \zeta)\big)_{1\le i,j\le 2}}{z_{1,2} + z_{2,1} - 2\operatorname{Tr} Z}, \qquad (10)$$

that is[6]:

$$\check{s}_\zeta(Z) = \begin{pmatrix} 2z_{1,1} + \zeta + \dfrac{(z_{1,1}+\zeta)^2}{z_{1,2}+z_{2,1}-2\operatorname{Tr} Z} & z_{1,2} - \dfrac{(z_{1,1}+\zeta)(z_{2,2}+\zeta)}{z_{1,2}+z_{2,1}-2\operatorname{Tr} Z} \\ z_{2,1} - \dfrac{(z_{1,1}+\zeta)(z_{2,2}+\zeta)}{z_{1,2}+z_{2,1}-2\operatorname{Tr} Z} & 2z_{2,2} + \zeta + \dfrac{(z_{2,2}+\zeta)^2}{z_{1,2}+z_{2,1}-2\operatorname{Tr} Z} \end{pmatrix}. \qquad (11)$$

The action of $\check{s}_\zeta$ on an unbalanced 2-port network impedance matrix $Z_0(s)$ is the impedance matrix $Z_1(s)$ of an unbalanced 2-port network made up of 3 copies of it, mutually linked according to the topology depicted in Fig. 6, using interconnection passive impedors of impedance $\zeta(s)$. In this case map $\check{s}_\zeta$ is the analog of map $T$ as defined in Subsection *IV.A*.

Iterating $\check{s}_\zeta$ map over and over produces the impedance matrices' sequence $(Z_n(s))_{n\in\mathbb{N}}$ of Sierpinski prefractal 2-port networks. A simpler model can be obtained by short-circuiting network's copies, i.e. letting $\zeta(s) \equiv 0$, as shown in Fig. 7.

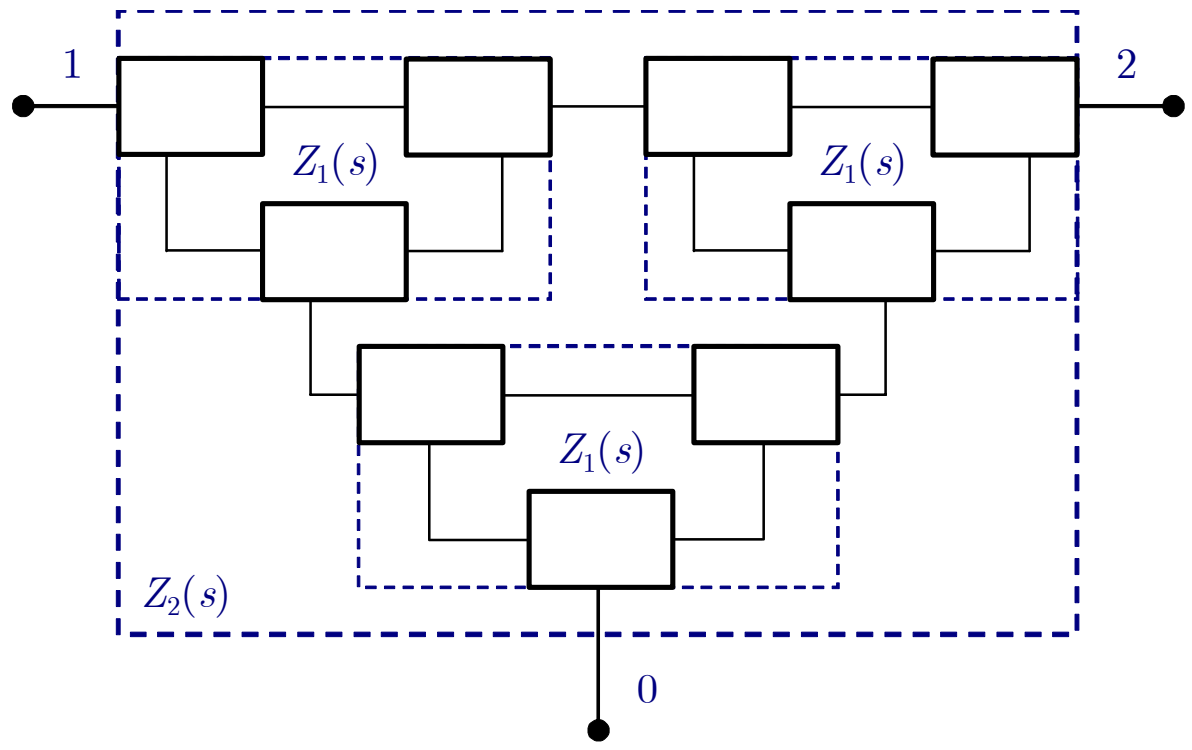

Fig. 7. Second step $Z_2 = \check{s}_0(\check{s}_0(Z_0))$ in the construction of the short-circuited Sierpinski 2-port network (without interconnection impedors).

Map $\check{s}_\zeta$, as a function of $Z \in \mathbb{M}_2(\mathbb{C})$ and $\zeta \in \mathbb{C}$, is *1-homogeneous*, i.e.

$$\check{s}_{\alpha\zeta}(\alpha Z) = \alpha \check{s}_\zeta(Z), \quad \forall \alpha \in \mathbb{C}. \qquad (12)$$

Singularity domain for maps $\check{s}_\zeta$ is the 1-dimensional locus of matrices $Z$ satisfying the condition

$$\operatorname{Tr} Z = \frac{z_{1,2} + z_{2,1}}{2}. \qquad (13)$$

These facts suggest that the proper Banach algebra for the map's analysis is the complex projective manifold $\mathbb{M}_2(\mathbb{C})/\mathbb{C}$, whose elements are the 1-dimensional spaces of *mutually proportional* $2\times 2$ complex matrices. An element $[\![Z]\!] \in \mathbb{M}_2(\mathbb{C})/\mathbb{C}$ is the equivalence class of proportional matrices to a sample matrix $Z$. Of course $[\![\alpha Z]\!] = [\![Z]\!]$, $\forall \alpha \in \mathbb{C}^*$.

For simplicity's sake, only the case $\zeta = 0$ will be considered. The gradient $\nabla \check{s}_0$ of $\check{s}_0$ matrix map is represented by the following 4th-order tensor ($O_n$ is the zero $n\times n$ matrix):

[6] 'Tr' is the *trace* operator of a square matrix; in this case: $\operatorname{Tr} Z = z_{1,1} + z_{2,2}$.

$$\begin{aligned}\nabla\check{s}_0 &= \left(\frac{\partial \check{s}_0}{\partial z_{i,j}}\right)_{h,k} = \begin{pmatrix} \frac{\partial}{\partial z_{1,1}} & \frac{\partial}{\partial z_{1,2}} \\ \frac{\partial}{\partial z_{2,1}} & \frac{\partial}{\partial z_{2,2}} \end{pmatrix} \check{s}_0 \\ &= \begin{pmatrix} \begin{pmatrix} 2 & 0 \\ 0 & 0 \end{pmatrix} & \begin{pmatrix} 0 & 1 \\ 0 & 0 \end{pmatrix} \\ \begin{pmatrix} 0 & 0 \\ 1 & 0 \end{pmatrix} & \begin{pmatrix} 0 & 0 \\ 0 & 2 \end{pmatrix} \end{pmatrix} + \\ &+ \frac{\begin{pmatrix} \begin{pmatrix} 2z_{1,1} & -z_{2,2} \\ -z_{2,2} & 0 \end{pmatrix} & O_2 \\ O_2 & \begin{pmatrix} 0 & -z_{1,1} \\ -z_{1,1} & 2z_{2,2} \end{pmatrix} \end{pmatrix}}{z_{1,2}+z_{2,1}-2\operatorname{Tr} Z} + \\ &+ \frac{\begin{pmatrix} \begin{pmatrix} 2z_{1,1}^2 & -2z_{1,1}z_{2,2} \\ -2z_{1,1}z_{2,2} & 2z_{2,2}^2 \end{pmatrix} & \begin{pmatrix} -z_{1,1}^2 & z_{1,1}z_{2,2} \\ z_{1,1}z_{2,2} & -z_{2,2}^2 \end{pmatrix} \\ \begin{pmatrix} -z_{1,1}^2 & z_{1,1}z_{2,2} \\ z_{1,1}z_{2,2} & -z_{2,2}^2 \end{pmatrix} & \begin{pmatrix} 2z_{1,1}^2 & -2z_{1,1}z_{2,2} \\ -2z_{1,1}z_{2,2} & 2z_{2,2}^2 \end{pmatrix} \end{pmatrix}}{\left(z_{1,2}+z_{2,1}-2\operatorname{Tr} Z\right)^2};\end{aligned} \tag{14}$$

its $(i,j,h,k)^{\text{th}}$ element, $1\le i,j,h,k\le 2$, being the derivative of $\check{s}_0(Z)$'s $(h,k)^{\text{th}}$ matrix element with respect to $z_{i,j}$ variable.

It is easily shown that first summand of (14), which is written in a more compact form as $(\delta_{i,h}\delta_{j,k}+\delta_{i,j}\delta_{j,h}\delta_{h,k})_{1\le i,j,h,k\le 2}$, is constant. Other two summands, depending on first and second power of the usual denominator $z_{1,2}+z_{2,1}-2\mathrm{Tr}Z$ respectively, are symmetric with respect to both $(i,j)$ and $(h,k)$ couples of indices; third (last) summand is also anti-symmetric[7] with respect to $(h,k)$ indices. The whole $4^{\text{th}}$-order tensor is trivially *projective*, i.e. $\nabla\check{s}_0(\alpha Z)=\nabla\check{s}_0(Z)$, $\forall\alpha\in\mathbb{C}^*$, so it is embedded in $\mathbb{M}_2(\mathbb{C})/\mathbb{C}$ too.

The linearization[8] of $\check{s}_0$ for a given perturbation $\Delta Z\in\mathbb{M}_2(\mathbb{C})$,

$$\check{s}_0(Z+\Delta Z) = \check{s}_0(Z) + \nabla\check{s}_0\Big|_Z : (Z+\Delta Z) + \mathrm{o}\left(\|\Delta Z\|^2\right),$$

is still 1-homogeneous with respect to $Z$ and leads to analogue linearization in the $\mathbb{M}_2(\mathbb{C})/\mathbb{C}$ space:

$$[\![\check{s}_0(Z+\Delta Z)]\!] \cong [\![\check{s}_0(Z)]\!] + \nabla\check{s}_0\Big|_Z : [\![\Delta Z]\!], \tag{15}$$

The iterated network construction, i.e. considering map $[\![Z_n]\!] \mapsto [\![Z_{n+1}]\!] = [\![\check{s}_0^{n+1}(Z_0)]\!] = [\![\check{s}_0(Z_n)]\!]$ for large $n$'s and expliciting the contraction products above, leads to the asymptotic dominance of diagonal elements, double with respect to non-diagonal ones in $\mathbb{M}_2(\mathbb{C})/\mathbb{C}$. For example the one between the tensor's constant part and $\Delta Z$ is[9]:

$$\begin{pmatrix} \begin{pmatrix} 2 & 0 \\ 0 & 0 \end{pmatrix} & \begin{pmatrix} 0 & 1 \\ 0 & 0 \end{pmatrix} \\ \begin{pmatrix} 0 & 0 \\ 1 & 0 \end{pmatrix} & \begin{pmatrix} 0 & 0 \\ 0 & 2 \end{pmatrix} \end{pmatrix} : \Delta Z = \Delta Z + \operatorname{diag}\Delta Z\,. \tag{16}$$

Similar considerations hold for the other two contraction products, with tensors' symmetric and anti-symmetric components leading to symmetric matrices[10] :

$$\lim_{n\gg 1} [\![\check{s}_0^{n+1}(Z)]\!] = \operatorname{sym}[\![\check{s}_0^{n}(Z) + \operatorname{diag}\check{s}_0^{n}(Z)]\!]\,. \tag{17}$$

This leads to the following limit in the $\mathbb{M}_2(\mathbb{C})/\mathbb{C}$ space:

$$\lim_n [\![\check{s}_0^{n}(Z)]\!] = \left[\!\!\left[\begin{pmatrix} 2 & 1 \\ 1 & 2 \end{pmatrix}\right]\!\!\right]. \tag{18}$$

It doesn't mean that map $\check{s}_0$ absolutely converges: its single matrix elements can diverge (which they actually do), but diagonal elements approach to the same value, which is double with respect to common value of anti-diagonal elements, for $n\to\infty$.

It can be numerically found that a $3/5$ ratio, rescaling the matrix at each new iteration, makes the map converge: in this sense the asymptotic properties of $\check{s}_0$ map is:

$$\lim_n \frac{3^n}{5^n}\check{s}_0^{n}(Z_0) = z_\infty(s)\begin{pmatrix} 2 & 1 \\ 1 & 2 \end{pmatrix}. \tag{19}$$

The impedance matrices associated to the Sierpinski network model get rescaled (which technically means using smaller and smaller networks) and the limit impedance matrix is that of a triangular (or "π") network whose three impedors all have impedance $z_\infty(s)$ and which only depend on the initiator matrix $Z_0(s)$. This also physically means that, no matter what kind of non-reciprocal starting network is chosen, the limit Sierpinski network is always reciprocal (as its limit impedance matrix is symmetric) and homogeneous (i.e. made up of identical impedors), regardless of the starting network. $z_\infty(s)$ impedance exists thanks to the $3/5$ rescaling factor, otherwise the matrix sequence would be diverging; cfr. Fig. 8.

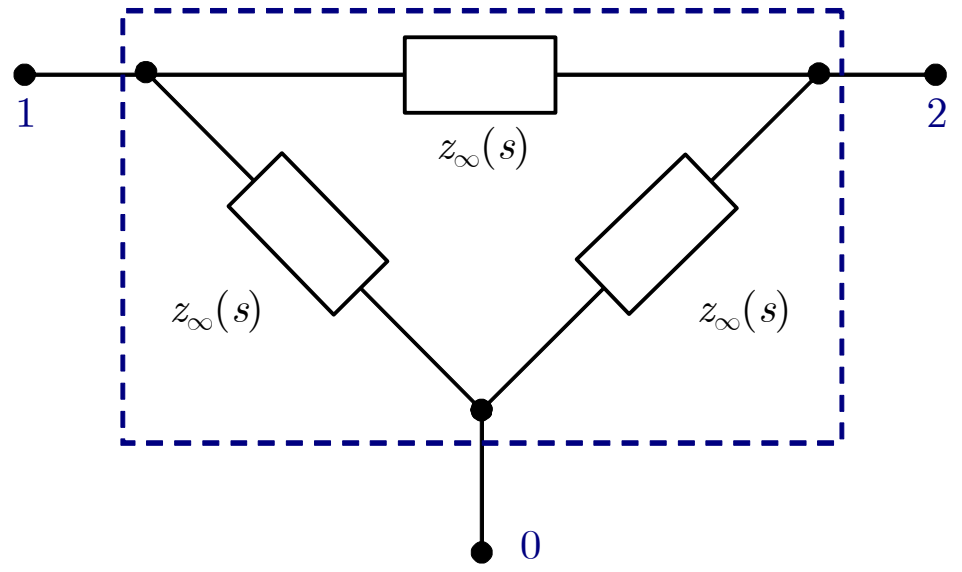

Fig. 8. Limit Sierpinski triangular network.

The same result can be proven as the limit matrix must be the fixed point for the $3/5$-rescaled Sierpinski network, thus leading to the following system of II-order equations:

$$\frac{3}{5}\check{s}_0(Z) = Z\,, \tag{20}$$

whose closed-form analytical solutions is:

$$Z = z_\infty\begin{pmatrix} 2 & 1 \\ 1 & 2 \end{pmatrix}, \tag{21}$$

with an undetermined scalar coefficient $z_\infty$.

[7] An *anti-symmetric* matrix is symmetric with respect to principal anti-diagonal.

[8] Symbol ':' denotes *contraction product* between tensors [15].

[9] $\operatorname{diag}A$ is the *diagonal part* of a square matrix $A$.

[10] $\operatorname{sym}A$ is the *symmetric part* of a matrix: $\operatorname{sym}A=(A+A^{\mathrm{T}})/2$.

### C. Numerical results and a "toy model"

Numerical simulation of the Sierpinski network model's convergence was performed. Random starting complex matrices (both symmetric and non-symmetric) were used. Figure 9 shows the values of the four impedance matrix elements versus the iteration order $n$. Convergence is always reached in around $30-40$ iterations, no matter what the matrix magnitude is, with diagonal elements relaxing on the same value, double of the common value for anti-diagonal elements.

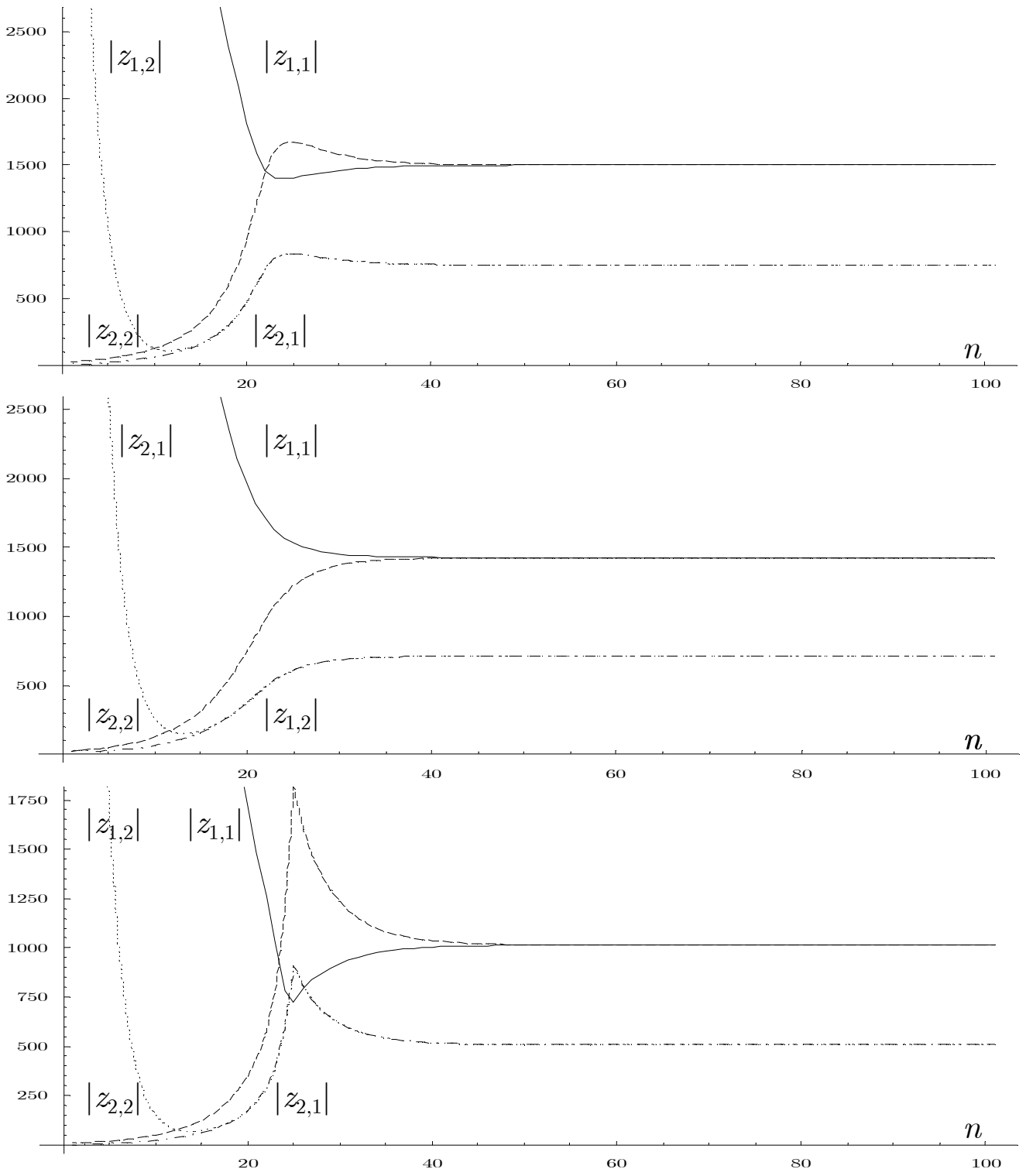

Fig. 9. Random examples of impedance matrix elements' moduli vs. iteration order $n$,

$0<n\leq 100$. Diagonal elements relax to the same value, which is double with respect to the anti-diagonal elements' one.

A "toy model" for this iterative network paradigm is presented at last to show some algebraic properties of $\check{s}_\zeta$ map. Let the starting unbalanced 2-port network be a homogeneous triangular one made with LC-parallel impedors and using resistor connections, i.e.:

$$Z_0(s) = \frac{1+s^2LC}{3sC}\begin{pmatrix} 2 & 1 \\ 1 & 2 \end{pmatrix}; \qquad \zeta(s) \equiv R\,. \qquad (22)$$

The wording "toy" is due to the chosen parameters' *normalization* for this example ($R$=1Ω, $L$=1H, $C$=1F), which is physically unfeasible. Introducing a rescaling, such impedance matrix can be made physically meaningful (and the graph shown in Fig. 10 is mathematically *similar* to a real one), but with these values some closed-form symbolic computations can be carried out. The dynamics of $\check{s}_R$ map is thus ($\forall n\in\mathbb{N}_0$):

$$Z_0(s) = \frac{1+s^2}{s}\begin{pmatrix} 2 & 1 \\ 1 & 2 \end{pmatrix}; \quad Z_{n+1}(s) = \check{s}_1\left(Z_n(s)\right) \quad (23)$$

The elements of symmetric matrix iterations $Z_n(s)$ are all rational functions with integer, symmetric coefficients and the same degrees for fixed $n$'s. Denominator polynomials are one degree higher than numerator ones: a simple pole in 0 is always present (as usual for LC immittances) and no private poles[11] are generated [15]. $Z_{n+1}(s)$ has twice the number of poles as $Z_n(s)$.

First two iterations $Z_1(s)$ and $Z_2(s)$ are computed here as an example:

$$Z_1(s) = \frac{\begin{pmatrix} N_{1,\mathrm{d}}(s) & N_{1,\mathrm{a}}(s) \\ N_{1,\mathrm{a}}(s) & N_{1,\mathrm{d}}(s) \end{pmatrix}}{15s(2s^2+s+2)}, \qquad (24)$$

$$\begin{cases} N_{1,\mathrm{d}}(s) = 2(s^2+1)(10s^2+9s+10) \\ N_{1,\mathrm{a}}(s) = 10s^4+15s^3+29s^2+15s+10 \end{cases};$$

$$Z_2(s) = \frac{\begin{pmatrix} N_{2,\mathrm{d}}(s) & N_{2,\mathrm{a}}(s) \\ N_{2,\mathrm{a}}(s) & N_{2,\mathrm{d}}(s) \end{pmatrix}}{75s(2s^2+s+2)\left(20s^4+24s^3+39s^2+24s+20\right)},$$

$$\begin{cases} N_{2,\mathrm{d}}(s) = 2(2s^2+3s+2)\cdot \\ \qquad \cdot(500s^6+480s^5+1437s^4+915s^3+1437s^2+480s+500) \\ N_{2,\mathrm{a}}(s) = 1000s^8+3540s^7+8494s^6+12663s^5+15186s^4+ \\ \qquad +12663s^3+8494s^2+3540s+1000 \end{cases}.$$

## V. Conclusion

With a full, lumped-parameters circuit model, as the "endoscopic model" presented in Section *III*, the voltage and current distributions give a detailed description of the electromagnetics within and the power distribution inside the network, once it is suitably tuned. As the number of nodes in a self-similar graph exponentially increases with $n$, the complexity to solve $\check{S}_n$'s electric circuit according to Subsection *III.A* is mainly due to the inversion of a $3^{n+1}\times 3^{n+1}$ admittances' matrix, symbolically depending on Laplace's variable $s\in\mathbb{C}$ (i.e., practically, from the signals' velocity $\omega\in\mathbb{R}$, for $s=i\omega\equiv 2\pi i\nu$).

This model was also shown to both independently agree with a full-wave MoM analysis [22] and a different scattering-matrix model [23] for Sierpinski gasket antennas.

On the other hand the "black-box" model presented in Chapter *IV* always produces a single multi-port network built from several copies of the previous iteration network. Model complexity is, *a priori*, always restricted to the fixed number of ports $N$ (in the case of the Sierpinski network, it's a $2\times 2$ impedance matrix) although only global parameters (i.e. referred to every port's degrees of freedom) can be usually extracted. These parameters are, on the contrary, the ones that are searched for whenever a circuit model is employed for the synthesis of planar or linear antennas, such as the *input impedance*.

[11] Any poles not possessed by immittance matrices' diagonal elements are called *private*: their presence may invalid a dynamical system's *inner* stability.

Of course the two methods coincide whenever the initiators are electrically the same (e.g. the triangle-shaped electric circuit $\check{S}_0$ is an unbalanced 2-port network with the ground node as unbalanced pin and using the same iterative paradigm), but once the "black-box" model is computed, antenna parameters are more easily extractable from the latter.

Such models were used to show the convergence properties of Sierpinski gasket-based networks which, in the high iteration order limit ($n \approx 40$, cfr. Subsection *IV.C*), always relax to a (reciprocal) triangular homogeneous network, which is arguable comparing the current distributions of Fig. 5 (at just the 9th iteration) with the limit structure of Fig. 9. This can be useful when synthesizing a simpler circuit model for a fractal antenna whose project constraints depend on the iteration (i.e. *pre*fractal) order: as circuit's complexity and cost increase with it, a proper evaluation of a required minimum iteration order $n$ is needed.

## Appendix

The purpose of this appendix is just to clarify the graph terminology and symbols used in Sections *II* and *III*.

In mathematics, a *graph* is an algebraic abstract object defined this way: let $N$ be a non-empty set (whose elements are called *nodes*), $L$ be any other set (even empty, whose elements are called *links*) and a relation $@\subseteq L\times N^2$ which is left-total and right-universal[12], called *incidence*. Such a 3-plet $G=(N,L,@)$ is called a *graph*. $G$ is said to be [in-]finite if so are both $N$ and $L$; *non-oriented* if $l@(m,n) \Leftrightarrow l@(n,m)$ (i.e. links are associated to a non-ordered couple of nodes); *connected* if $\forall m\ \exists n,\ \exists l$: $l@(m,n)$ (i.e. every node is associated at least one link); *simple* if '@' is 1-way (i.e. there cannot be more than one link per couple of nodes). $G$ is *complete* if it is connected and simple at the same time (i.e. every couple of nodes are incident to each other).

The case of non-oriented graphs is much simpler because the link-set can be roughly identified with a subset of nodes' couples, i.e. $@\cong L\subseteq N^2$. Incidence '@' becomes an equivalence relation and $G=(N,L)$. For a simple, non-oriented graph to be connected, $\mathrm{Card}L > \mathrm{Card}N$ condition is required, whereas $\mathrm{Card}L = (\mathrm{Card}N)(\mathrm{Card}N-1)/2$ is mandatory for completeness.

A graph with no closed paths (or *cycle*s, or *circuit*s in mathematics) is called a *tree*. Let $G=(N,L)$ be a finite, connected, non-oriented graph; any trees whose node-set is $N$ and whose link-sets $T_L$ are generated by adding links of $L$ one after another, such that the resulting graph is still a connected tree, are called *fundamental tree*s of $G$. Their complement graphs $(N,L\backslash T_L)$ are called the fundamental *co*trees of $G$, for every possible $T_L$.

To every finite, simple graph $G=(N,L,@)$ a skew-symmetric matrix $H_G \in \mathbb{M}_{\mathrm{Card}N}(\mathbb{Z}_3)$ is associated such that, once a node-ordering is chosen, its $(i,j)^{\mathrm{th}}$ element is either $\pm 1$ if $i^{\mathrm{th}}$ node is connected toward $j^{\mathrm{th}}$ node [or vice versa], or 0 if such a link does not exist. Such matrices are called *incidence matrices*, and there exists a complete isomorphism between them and such graphs, such that algebraic properties of the former reflect topological properties of the latter.

Non-oriented graphs have symmetric, zero-diagonal incidence matrices instead, with elements in $\mathbb{Z}_2 \cong \{0,1\}$: their $(i,j)^{\mathrm{th}}$ element being either 1 or 0 according to whether $i^{\mathrm{th}}$ and $j^{\mathrm{th}}$ node are linked or not, respectively.

[12] *Left-total* means that $\forall l\in L\ \exists n',n''\in N$: $l@(n',n'')$; *Right-universal* means that $l@(n',n'')$, $l@(m',m'')$, $n'\neq m'' \Rightarrow (n',n'')=(m',m'')$. That is, each link is associated to no more than one *orderless* couple of nodes.